# The microscopic theory of the upper critical field Hc2 in two-band systems in the magnetic field directed parallel and perpendicular to the plane (ab). Application to MgB2

M.E.Palistrant, I.D. Cebotari, V.A. Ursu,
*Institute of Applied Physics*
*Academiei str., 5, 2028, Chisinau, Moldova*

## Abstract

The microscopic superconductivity theory in $MgB_2$ systems in the magnetic fields directed parallel and perpendicular to the plane (ab) is built. The two energy bands of different dimensions overlapping on the Fermi surface (two-dimensional $\sigma$ - band and three-dimensional $\pi$ -band) are taken into account. The theory is built taking into account the weak dispersion of the electron energy of $\sigma$ -band in z direction.

The systems of equations for the superconductive transition temperature $T_c$ and the upper critical field $H_{c2}(ab)$ are obtained. The analytic solutions of the equations for $H_{c2}^{(ab)0}$ are found in the two temperature areas: nearby the zero temperature $T \ll T_c^0$ and in the vicinity of the superconductive transition temperature $T_c^0 - T \ll T_c^0$ .

The temperature dependencies of the upper critical felds $\vec{H}_{c2}(ab)$ and $\vec{H}_{c2} \| c$ on the temperature are built, and the anisotropy coefficient $\gamma_H = H_{c2}^0 \|(ab)/ H_{c2}^0 \| c$ is determined. It is shown that the value of $H_{c2}(ab)$ exceeds greatly the value of $H_{c2}^0(c)$ due to the system anisotropy (to small values of the average velocity of electrons in the crystallographic axis direction).

## 1. Introduction

Superconductivity in the intermetallic compound MgB2 at temperature $T_c \sim 40K$ was discovered by a group of Japan researchers [1] in the year 2001. This discovery aroused a great interest because of relatively simple crystal structure of MgB2, high $T_c$ values, as well as large values of critical magnetic fields and critical currents that let consider this material as a high-temperature compound. Investigations had shown that it wasn't possible to describe the high-temperature properties of MgB2 by using the theory of superconductivity of BKS-Bogoliubov-Eliashberg, which had been created for the isotropic systems. The MgB2 compound is an anisotropic system and its anisotropy becomes apparent mainly in band structure. According to the band structure of MgB2 [2], several energy bands are intersecting on the Fermi surface of this material. Thus two energy bands of different dimension play the decisive role in the matter of superconductivity: two-dimensional σ - band and three-dimensional π-band (hereinafter we'll term them band 1 and band 2 correspondingly).

In a number of researches (e. g. [3], [4]) it was recommended to apply the two-band Moskalenko model [5] (see also H. Suhl et al [6]), which had been suggested in the year 1959, to describe superconductive properties of MgB2.

The articles [5] and [6] were the first works, which let to take into consideration the properties of the real superconductors. In particular, there where shown the violation of universality of a series of physical quantities relations, which took place in the isotropic one-band superconductive systems. It is necessary to note that long before the discovery of high-temperature superconductivity, all the more the discovery of superconductivity in MgB2, on the basis of the model [5], [6] it was built the theory of thermodynamic, magnetic and kinetic properties (pure and impure) with overlapping on the Fermi surface energy bands. Let's mark that in that distant past the scientists of many countries made contribution in the development of the two-band superconductor theory, but for all that the most important contribution belongs to the school of the Moldavian physicists with V.A. Moskalenko at the head.

The history of development of the multi-band superconductors theory can be found particularly in works [7], [8], as well as in the review articles [9], [10] and books [11-13].

The discovery of superconductivity in $MgB_2$ and the experimental confirmation of existing of two energy gabs awoke a great interest and stimulated an intensive investigation of its properties. A lot of works were written, unfortunately, many of them had repeated the results of Moldavian scientists of many years remoteness. Along with this in a great number of investigations it was obtained the valuable information about $MgB_2$ properties: a phonon mechanism of superconductivity had been confirmed, it had been revealed the crystalline structure, the original band structure, which had corresponded to the overlapping of energy bands on the Fermi surface of different topology, the original role of every of these bands and a number of other properties peculiar to $MgB_2$ compound (e. g. [14]-[16]).

In compliance with the classic results ([9], [12]) the two-band model [5] lets to obtain high values of the superconductivity transition temperature $T_c$, two energetic gabs $\Delta_1$ and $\Delta_2$ (at the same time the conditions $2\Delta_1/T_c > 3,5$ and $2\Delta_2/T_c < 3,5$ hold true), low values of the relative jump of electronic thermal capacity in the point $T = T_c$ (less than 1,43 – the value which is typical for one-band case, in two-band $MgB_2$ the value is 0,8), anomalous temperature dependence of thermal capacity in the superconductivity phase, positive curvature of the upper critical field $H_{c2}$ near the temperature of the superconductivity transition, violation of the Anderson's theorem in the superconductors with nonmagnetic impurity etc. A lot of the experimental researches of the $MgB_2$ superconductive properties confirm above-stated theoretical theses.

The theory based on the two-band model describes well the thermodynamical properties of either pure $MgB_2$, or the properties of the systems where the atoms of Mg and B are substituted by the other elements of the periodic table of the elements (e. g. [17], [18]). Along with this $MgB_2$ is a strong anisotropic system and has a number of additional features in comparison with an isotropic one-band system. Such features in particularly are the relative location and small width of the energetic bands, also the reduced dimension of one of them. Every of these bands (two-dimensional $\sigma$-band and three-dimensional $\pi$ - band) plays its own role in the question of the appearance of superconductivity and revealing of the anomaly of some physical quantities in $MgB_2$.

The experimental investigations of the $MgB_2$ magnetic properties show bright appearance of anisotropy of the upper critical field $H_{c2}$ [19]. The upper critical field $H_{c2}^{(ab)}$, which corresponds to the external magnetic field in the plane ($ab$), exceeds in a few times the $H_{c2}^{(c)}$ value, which responds to the parallel with the $c$-axis magnetic field. The theoretical researches of pure $MgB_2$ in the magnetic field (close to the upper critical field) on the basis of the two-band model in a quasi-classic approximation were accomplished in the works [20] and [21].

We set a problem to build the microscopic theory of the upper critical field $H_{c2}$ of a pure anisotropic two-band superconductor, applicable on the whole temperature interval $0 < T < T_c$, and to describe the pattern of the $H_{c2}$ value behavior as a function of temperature in $MgB_2$, also to determine the curvature of the upper critical field $H_{c2}^{(ab)}$ and $H_{c2}^{(c)}$ close to the temperature of superconducting transition, to reveal then the anisotropy of temperature dependence of the coefficient $\gamma_H = H_{c2}^{ab} / H_{c2}$ . Also we'll try to determine the influence of the energy bands occupation mechanism on the values $T_c$ and $H_{c2}^{(ab)}$ when the system is doped with electrons or holes.

In the base of the two-band systems researches lies the microscopic approximation of the theory of superconductivity [22] – [24].

The work is presented in the following way. In section 2 is adduced the system of Ginzburg-Landau equations for a pure two-band superconductor in the magnetic field $H^{(ab)}$ close to the value of the upper critical field $H_{c2}^{(ab)}$. The investigation of the two-band system properties in the magnetic field is accomplished on the basis of generalization of the Maki and Tsuzuki method, which was developed in one's time for a one-band superconductor [23] and some time later by one of the authors for a two-band isotropic system [25] (see also [10]). In this section the result is adduced as in [9], some transformations and calculations are accomplished in the system of equations for the order parameters $\Delta_m^*$ a two-band superconductivity system (m = 1, 2), which let then to obtain the equations for the superconductivity transition temperature $T_c(\tilde{n})$ and also the equation for determination of the upper critical field $H_{c2}^{(ab)}(\tilde{n})$ in a system with variable density of the charge carriers.

In section 3 are adduced the equations for determination of $H_{c2}^{(ab)}$ either for a pure two-band system, or for the case when the system is doped with electrons or holes. Section 4 is dedicated to the determination of the analytic solutions of the equation for $H_{c2}^{(ab)}$ on the basis of the asymptotic values of the occurred in the equation functions close to zero temperature and close to the temperature of superconductive transition.

In section 5 the values of the upper critical field $H_{c2}$ close to superconductive transition temperature $(T_c - T \ll T_c)$ and nearby zero temperature $T \ll T_c$ are adduced.

In section 5 the numerical computation and connected with the system anisotropy discussions of the results are given.

## 2. The system of equations for the order parameters $\Delta_m^*$ in the magnetic field $\vec{H}$ directed parallel to the plane *(ab)*

We consider high values of the magnetic field $H$ (close to the value of upper critical field $H_{c2}$), i.e. the subcritical region nearby unstable normal state.

In this region the system of equations for the order parameters $\Delta_m^*$ of a two-band superconductor (m = 1,2) is given by

$$\Delta_m^*(\vec{r}) = \frac{1}{\beta} \sum_\omega \sum_n V_{nm} \int d\vec{r}\,' g_n(\vec{r}\,',\vec{r},\omega) \Delta_n^*(\vec{r}\,') g_n(\vec{r}\,',\vec{r},-\omega) -$$
$$- \frac{1}{\beta} \sum_\omega \sum_n V_{nm} \int d\vec{r}_1 d\vec{r}_2 d\vec{r}_3 \Delta_n^*(\vec{r}_1) \cdot g_n(\vec{r}_1,\vec{r}_2,-\omega) \Delta_n^*(\vec{r}_3) g_n(\vec{r}_3,\vec{r}_1,\omega) g_n(\vec{r}_3,\vec{r},-\omega) \qquad (1)$$

Where $V_{mn}$ is intra-band (n=m) and inter-band $n \neq m$ electron- electron interactions. Electron Green function $g_n$ in presence of a magnetic field is defined by expression [22]:

$$g_n(\vec{r},\vec{r}\,',\omega) = e^{i\varphi(\vec{r}\,\vec{r}\,')} g_n^0(r,r',\omega) \qquad (2)$$

Here $g_n^0$ is a Green function in normal state without magnetic field. The presence of a magnetic field is taken into consideration by the phase multiplier:

$$\varphi(\vec{r}\,',\vec{r}) = \int_{\vec{r}}^{\vec{r}\,'} A(\vec{l})\,d\vec{l}. \qquad (3)$$

The system of equations (1) was obtained on the basis of the results of work [24] in the diagonal approximation by the band's indexes. Such approximation takes into account the processes of superconducting pairing of electrons into every energy band and their tunneling as a whole from one band into another.

At $\vec{H} \to \vec{H}_0 = \vec{H}_{c2}$ the order parameters $\Delta_m \to 0$, and it is possible to confine by the linear values of $\Delta_n$ terms in the equation (1). With all this we obtain:

$$\Delta_m^*(\vec{r}) = \frac{1}{\beta} \sum_n V_{mn} \sum_\omega \int d\vec{r}\,' g_n(\vec{r}\,',\vec{r},\omega) \Delta_n^*(\vec{r}\,') g_n(\vec{r}\,',\vec{r},-\omega). \qquad (4)$$

We'll substitute in this formula definition (2), performing expansion of the function $g_n^0$ by the Bloch functions, then we'll average this equation over the amplitudes of the Bloch functions. By this way we have:

$$\Delta_m^*(\vec{r}) \quad \sum_n V_{mn} \frac{1}{\beta} \sum_\omega \int d\vec{r}\,' \sum_{\vec{k}} \sum_{\vec{q}} g_n^0(\vec{k},\omega)\, g_n^0(\vec{q}-\vec{k},-\omega) \Delta_n^*(\vec{r}\,') e^{2i\varphi(\vec{r}\,',\vec{r})} e^{i\vec{q}(\vec{r}\,'-\vec{r})} \qquad (5)$$

$$g_n^0(\vec{k},\omega) = [\,i\,\omega - \xi_n(\vec{k})\,]^{-1}, \qquad (6)$$

where $\omega = \pi T(2n+1)$ is the Matsubar frequency, $\xi_n$ – the electron energy in the n$^{th}$ band.

Let's consider the magnetic field $\vec{H}$ (which is parallel to the plane $(ab)$) directed along the y-axis. At that it is possible to choose $A_x = -\dfrac{H_0}{2}(x+x') = A_y = A_z \quad 0$ in the symmetric view and on the basis of (3) we obtain:

$$2\varphi(\vec{r}\,',\vec{r}) = e\,H_0(x+x')(z-z'). \qquad (7)$$

We'll represent the dispersion law for $\sigma$- and $\pi$ - bands (it is marked by index 1 and 2 correspondingly) in the form:

$$\xi_1(\vec{k}) = \zeta_1 + \frac{k_x^2 + k_y^2}{2m_1} + \frac{k_z^2}{2M} - \mu \quad \text{и} \quad \xi_2(\vec{k}) = \zeta_2 + \frac{k_x^2 + k_y^2 + k_z^2}{2m_2} - \mu \;, \qquad (8)$$

where $M \gg m_1$. Satisfiability of this inequality means weak deviation of the dispersion law of the first band from two-dimensionality. The Fermi surface of the second band we suppose spherical for simplicity.

We'll introduce the definition of function $g_n^0$ (6) in (5) and we'll pass from summation by $\vec{k}$ to integration by energy in every energy band in accordance with the dispersion law (8) (by the cylindrical space in band 1 and the spherical space in band 2). As a result of these operations the equation (5) takes form

$$\Delta_1^*(\vec{r}) \quad -V_{m1} \sum_\omega \frac{m_1}{(2\pi)^3} \int d\vec{r}\,' \Delta_1^*(\vec{r}\,') e^{i\vec{q}(\vec{r}-\vec{r}\,')+ieH_0(x+x')(z-z')} \int\limits_{-P_0}^{P_0} dp_z$$

$$\int\limits_0^{2\pi} d\varphi \int\limits_{-\infty}^{\infty} d\xi_1^0 \frac{1}{[i\omega - \xi_1^0(p_1) - \varepsilon \frac{p_z^2}{2m_1}]} \left[ i\omega + \xi_1^0(p_1) + \frac{\varepsilon\, p_z^2}{2m_1} + v_x q_x + v_y q_y + \varepsilon \frac{p_z}{m_1} q_z \right] -$$

$$- V_{m2} \sum_\omega \int a\vec{r}\,' \Delta_2^*(\vec{r}\,') \frac{m_2\, p_{F_2}}{2\pi^3} \int\limits_0^{2\pi} d\varphi \int\limits_0^{\pi} \sin\theta\, d\theta \int \frac{dq}{(2\pi)^3} \int\limits_{-\infty}^{\infty} d\xi_2 \times$$

$$\times \frac{\exp[i\vec{q}(\vec{r}-\vec{r}\,') + ieH_0(x+x')(z-z')]}{(i\omega - \xi_2)(i\omega + \xi_2 + v_{2x} q_x + v_{2y} q_y + v_{2z} q_z)} \qquad (9)$$

where

$$\xi_1^0(p) = \zeta_1 + \frac{p_x^2 + p_y^2}{2m_1} - \mu, \quad \varepsilon = \frac{m_1}{M} \ll 1. \qquad (10)$$

$p_0$ - is an impulse of the cylinder cut off along the axis $z$, $v_{ix}$, $v_{iy}$, $v_{iz}$ – are corresponding projections of the electron velocity on the i$^{th}$ cavity of the Fermi surface. In the first band $v_{1x} = v_1 \cos\varphi$, $v_{1y} = v_1 \sin\varphi$ in the second band $v_{2x} = v_2 \cos\theta$, $v_{2y} = v_2 \sin\theta \cos\varphi$, $v_{2z} = v_2 \sin\theta \sin\varphi$

We'll consider the order parameters $\Delta_m^*(\vec{r})$ depend on $x$ [22], [23]. Taking into account this depending we'll perform a series of operations in the system of equations (9): integration by energy $\xi_1^0$ and $\xi_2$, partial integration by coordinate variables and by $\vec{q}$, summation by $\omega$. By this way we obtain

(m = 1, 2):

$$\Delta_1^*(x) = \lambda_{11} \frac{2T}{v_1} \int_1^\infty \frac{du}{\sqrt{u^2-1}} \int_{-\infty}^\infty \frac{\Delta_1^*(x')\,dx'}{\sinh \dfrac{2\pi T\,|x-x'|u}{v_1}} \frac{\sin(\tilde\varepsilon eH_0 u(x^2-x'^2))}{\tilde\varepsilon eH_0 u(x^2-x'^2)} \theta(|x-x'|-\delta_{11}) +$$

$$+ V_{12}N_2 \frac{\pi T}{v_2} \int_1^\infty \frac{du}{u} \int_{-\infty}^\infty \frac{dx'\Delta_2^*(x')}{\sinh[\dfrac{2|x-x'|\pi Tu}{v_2}]} J_0\left[(x^2-x'^2)eH_0\sqrt{u^2-1}\right]\theta\left(|x-x'|-\delta_{12}\right)\right), \tag{11}$$

$$\Delta_2^*(x) = \lambda_{21} \frac{2T}{v_1} \int_1^\infty \frac{du}{\sqrt{u^2-1}} \int_{-\infty}^\infty \frac{\Delta_1^*(x')\,dx'}{\sinh \dfrac{2\pi T\,|x-x'|u}{v_1}} \frac{\sin(\tilde\varepsilon eH_0 u(x^2-x'^2))}{\tilde\varepsilon eH_0 u(x^2-x'^2)} \theta(|x-x'|-\delta_{21}) +$$

$$+ V_{22}N_2 \frac{\pi T}{v_2} \int_1^\infty \frac{du}{u} \int_{-\infty}^\infty \frac{dx'\Delta_2^*(x')}{\sinh[\dfrac{2|x-x'|\pi Tu}{v_2}]} J_0\left[(x^2-x'^2)eH_0\sqrt{u^2-1}\right]\theta\left(|x-x'|-\delta_{22}\right), \tag{12}$$

where $\tilde\varepsilon = \dfrac{p_0\,\varepsilon}{m_1 v_1}$ .

Every term of the right part of the system of equations (11), (12) contains step function $\theta(|x-x'|)-\delta_{ij})$, which responds to the cut off of the integrals in the impulse space by definition of the parameters $\delta_{ij}$ [23].

Along with aforesaid operations in the system of equations (11) and (12) it were introduced renormalized constants of the two-band theory related with the strong electron-phonon coupling and Coulomb electron-electron interaction. As a result constants $\lambda_{mn}$ take form [17]:

$$\lambda_{11} = \frac{\lambda_{11}^0 - \mu_{11}^*}{1+\lambda_{11}^0+\lambda_{12}^0}, \ \lambda_{12} = \frac{\lambda_{12}^0 - \mu_{12}^*}{1+\lambda_{11}^0+\lambda_{12}^0},$$

$$\lambda_{21} = \frac{\lambda_{21}^0 - \mu_{21}^*}{1+\lambda_{22}^0+\lambda_{21}^0}, \ \lambda_{22} = \frac{\lambda_{22}^0 - \mu_{22}^*}{1+\lambda_{22}^0+\lambda_{21}^0}. \tag{13}$$

Here $\lambda_{mn}^0 = V_{mn}N_n$, where $N_n$ is the electronic states density in the n$^{th}$ energy band $\left(N_1 = \dfrac{m_1 p_0}{2\pi^2}, N_2 = \dfrac{m_2 p_{F2}}{2\pi^2}\right)$, $\mu_{mn}^*$ - Coulomb electron-electron interactions.

If the inter-band interaction is neglected, what can be done when $\lambda_{11}, \lambda_{22} \ll \lambda_{12}, \lambda_{21}$, then on the basis of (11) and (12) we'll obtain two independent equations, form which it follows that at $\tilde\varepsilon \to 0$ $\Delta_1^*(x)$ doesn't depend on the magnetic field value. This fact we'll use at choosing the solutions for the order parameters $\Delta_1^*(x)$ and $\Delta_2^*(x)$ conformably to MgB$_2$, inasmuch as inter-band electron-phonon interactions are low for this compound. This circumstance allows choosing solution for parameters $\Delta_1^*(x)$ and $\Delta_2^*(x)$ in the form:

$$\Delta_1^*(x) = \Delta_1^* e^{-\tilde\varepsilon eH_0 x^2}, \ \Delta_2^*(x) = e^{-eH_0 x^2} \tag{14}$$

In that way $\lim_{\varepsilon\to 0}\Delta_1^*(x) \quad \Delta_1^*$.

We'll put the values (14) into the system of equations (11) and (12), then we'll multiply the equation (11) on $e^{-\tilde\varepsilon eH_0 x^2}$, and (12) on $e^{-eH_0 x^2}$, after that we'll perform in the both equations integration by variable $x$ in the infinite limits. Then we'll bring the system of equations (11) and (12) passing on to dimensionless variables like $\xi = (\tilde\varepsilon eH_0)^{1/2} x$ and $\xi' = (\tilde\varepsilon eH_0)^{1/2} x'$ and having performed a series of other transformations (see appendix A) to the form:

$$\Delta_1^* = \Delta_1^* \lambda_{11} \xi^{(1)}(T_c) + \Delta_1^* \lambda_{11} [\xi^{(1)}(T) - \xi^{(1)}(T_c)] - \lambda_{11} f_{11}(\rho_1 \tilde{\varepsilon}) \Delta_1^* + \tilde{\lambda}_{12} \Delta_2^* \xi^{(2)}(T_c) + \tilde{\lambda}_{12} [\xi^{(2)}(T) - \xi^{(1)}(T_c)] \Delta_2^* - \tilde{\lambda}_{12} f_{12}(p_2 \tilde{\varepsilon}) \Delta_2^*,$$ (15)

$$\Delta_2^* = \tilde{\lambda}_{21} \Delta_1^* \xi^{(1)}(T_c) + \tilde{\lambda}_{21} \Delta_1^* [\xi^{(1)}(T) - \xi^{(1)}(T_c)] - \tilde{\lambda}_{21} f_{21}(\rho_1 \tilde{\varepsilon}) \Delta_1^* + \lambda_{22} \Delta_2^* \xi^{(2)}(T_c) + \lambda_{22} [\xi^{(2)}(T) - \xi^{(2)}(T_c)] - \lambda_{22} f_{22}(\rho_2 \tilde{\varepsilon}) \Delta_2^*,$$ (16)

where

$$\xi^{(n)}(T) = \int_{-d_n}^{d_{cn}} d\varepsilon \frac{th \frac{\beta \varepsilon}{2}}{2\varepsilon}, \ \xi^{(n)}(T_c) = \int_{-d_n}^{d_{cn}} d\varepsilon \frac{th \frac{\beta_c \varepsilon}{2}}{2\varepsilon}, \ \tilde{\lambda}_{12} = \tilde{\varepsilon}^{1/2} \sqrt{\frac{2}{\tilde{\varepsilon}+1}} \lambda_{12}, \ \tilde{\lambda}_{21} = \sqrt{\frac{2}{\tilde{\varepsilon}+1}} \lambda_{21},$$ (17)

$\beta = 1/T$, and values $d_n = \mu - \zeta_n$, $d_{cn} = \zeta_{cn} - \mu$ are the parameters of the integrals cut off by energy at variable density of the charge carriers. Under phonon mechanism of superconductivity (the MgB$_2$ case) these parameters have form:

$$d_n = \begin{cases} \mu - \zeta_n & \text{at} \quad \mu - \zeta_n \leq \omega_n, \\ \omega_n & \text{at} \quad \mu - \zeta_n > \omega_n, \end{cases}$$

$$d_{cn} = \begin{cases} \omega_n & \text{at} \quad \zeta_{cn} - \mu > \omega_n, \\ \zeta_{cn} - \mu & \text{at} \quad \zeta_{cn} - \mu < \omega_n, \end{cases}$$ (18)

Here $\omega_n$ is the characteristic phonon frequency corresponding to the n[th] energetic band. The functions $f_{nm}$ which contain the dependence on the magnetic field have the appearance:

$$f_{11} = \frac{(\tilde{\varepsilon}\rho_1)^{-1/2}}{\pi} \int_{-1}^{1} dy \int_{1}^{\infty} \frac{du}{\sqrt{u^2-1}} \int_{0}^{\infty} \frac{d\zeta}{\sinh \frac{\zeta u}{(\tilde{\varepsilon}\rho_1)^{1/2}}} (1 - \exp(-\frac{\zeta^2}{2}(1+u^2 y^2))),$$ (19)

$$f_{12} = \rho_2^{-1/2} \int_{0}^{\pi} d\varphi \int_{1}^{\infty} \frac{du}{u} \int_{0}^{\infty} \frac{d\zeta}{\sinh(\zeta u / \rho_2^{1/2})} \left( 1 - \exp\left( \frac{-\zeta^2(1+\tilde{\varepsilon})}{4} \left[ 1 - \left( \frac{\tilde{\varepsilon} - 1 + 2i\sqrt{u^2-1} \cos \varphi}{1+\tilde{\varepsilon}} \right)^2 \right] \right) \right),$$ (20)

$$f_{21} = \frac{\rho_1^{-1/2}}{\pi} \int_{-1}^{1} dy \int_{1}^{\infty} \frac{du}{\sqrt{u^2-1}} \int_{0}^{\infty} \frac{d\zeta}{\sinh \frac{\zeta u}{\rho_1^{1/2}}} \left( 1 - e^{\frac{-\zeta^2(1+\tilde{\varepsilon})}{4} \left[ 1 - \left( \frac{1-\tilde{\varepsilon}-2i\tilde{\varepsilon}uy}{1+\varepsilon} \right)^2 \right]} \right),$$ (21)

$$f_{22} = \rho_2^{-1/2} \int_{1}^{\infty} \frac{du}{u} \int_{0}^{\infty} \frac{d\zeta}{\sinh \frac{\zeta u}{\rho_2^{1/2}}} \left[ 1 - e^{-\frac{\zeta^2}{4}(u^2+1)} I_0 \left( \frac{\zeta^2(u^2-1)}{4} \right) \right],.$$ (22)

In the definition of functions $f_{nm}$ was introduced the dimensionless parameter $\rho_n^{1/2} = \frac{v_n(eH_0)^{1/2}}{2\pi T}$, which contains the value $v_n$ is the velocities of electrons on n[th] the cavity of the Fermi surface.

## 3. The equations for determination of the upper critical field H$_{c2}$ and critical temperature of the superconductive transition T$_c$

We'll supplement the system of equations (15) and (16), which is the basis for determination of the upper critical field $H_0 = H_{c2}^{(ab)}$ with the system of equations for determination of the superconductivity transition temperature $T_c$. In the limit $H \to \infty$ the system of equations for the order parameters $\Delta_{n0}^*$ has the appearance:

$$\Delta_{10}^* = \tilde{\lambda}_{11} \Delta_{10}^* \xi^{(1)}(T) + \tilde{\lambda}_{12} \Delta_{20}^* \xi^{(2)}(T),$$
$$\Delta_{20}^* = \tilde{\lambda}_{21} \Delta_{10}^* \xi^{(1)}(T) + \lambda_{22} \Delta_{20}^* \xi^{(2)}(T), \tag{23}$$

From the condition of solvability of this system for the superconductivity transition temperature $\quad T = T_c$ we obtain:

$$a \, \xi^{(1)}(T_c) \xi^{(2)}(T_c) - \lambda_{11} \, \xi^{(1)}(T_c) - \lambda_{22} \, \xi^{(2)}(T_c) + 1 = 0, \tag{24}$$

where $a = \lambda_{11}\lambda_{22} - \tilde{\lambda}_{12}\tilde{\lambda}_{21}$, and $\xi^{(n)}(T)$ is given by the formula (17). Equating with zero determinant of the system (14), (16) and using the equation (24), we obtain the equation for determination of the upper critical field $H_0 = H_{c2}^{(ab)}$, with which the superconductive pairs appear. This equation has form:

$$\lambda_{11} \lambda_{22} \tilde{F}_{11} \tilde{F}_{22} - \tilde{\lambda}_{12} \tilde{\lambda}_{21} \tilde{F}_{12} \tilde{F}_{21} + \lambda_{11} \left[ 1 - \lambda_{22} \, \xi^{(2)}(T_c) \right] \tilde{F}_{11} + \lambda_{22} \left[ 1 - \lambda_{11} \, \xi^{(1)}(T_c) \right] \times$$
$$\times \tilde{F}_{22} + \tilde{\lambda}_{12} \tilde{\lambda}_{21} \, \xi^{(2)}(T_c) \, \tilde{F}_{21} + \tilde{\lambda}_{12} \tilde{\lambda}_{21} \, \xi^{(1)}(T_c) \, \tilde{F}_{12} = 0, \tag{25}$$

where

$$\tilde{F}_{mn} = f_{mn} + \xi^{(n)}(T_c) - \xi^{(n)}(T), \tag{26}$$

$$f_{mn} = f_{mn}(\rho_n, \tilde{\varepsilon}), \qquad \xi^{(n)}(T_c) - \xi^{(n)}(T) = -\ln T_c / T. \tag{27}$$

For the investigating of the superconductive properties of the system with variable density of the charge carriers it is necessary to supplement the equation (25) with the condition which define the chemic potential $\mu$ (the charge carriers density $\tilde{n}$) [26, 27]:

$$\tilde{n} = \sum_m N_m \int_{-d_m}^{d_{cm}} d\varepsilon_m \left[ \frac{E_m(n) - |\varepsilon(\vec{k}) - \mu|}{E_m(\vec{k})} + \frac{2|\varepsilon_m(\vec{k}) - \mu|}{E_m(\vec{k})} \frac{1}{1 + \exp\beta \, E_m(\vec{k})} \right]. \tag{28}$$

Considering $d_m / T_c$ and $d_{mc} / T_c \gg T_c$ on the basis of the correlation (29) we obtain [17]:

$$\tilde{n} = \sum_n Nn[\zeta_{cn} - \zeta_n - |\zeta_{cn} - \mu| + |\zeta_n - \mu|]. \tag{29}$$

The system of equations (24), (25) and (29) allows us to determine the value of the critical temperature $T_c$ and upper critical field $H_{c2}^{(ab)}(T)$ at any density of the charge carriers. If we set the chemical potential at the level of its value for pure MgB$_2$ ($\mu = \mu_0$ in this case), then the above system of equations allow us to determine the temperature of the superconductive transition $T_c^0$ and the upper critical field $H_{c2}^{(ab)0}$ of pure MgB$_2$.

The analysis will be done in this paper, and also the results for the case of pure $MgB_2$ are obtained. The influence of the dopping will be considered later.

## 4. Determination of the upper critical field H$_{c2}$ ∥ (ab). The analytical solutions

The equations (25) contain difficult integral dependencies $f_{mn}$ (19) – (22), it is possible to solve such equation on the whole temperature interval $0 < T < T_c^0$ only by the numerical methods. However, it is possible to find the analytical solutions of this equation in two limit cases: nearby the critical temperature $\tilde{\varepsilon}\rho_n \ll 1$, when $(T_c^0 - T \ll T_c^0)$ and in the area of low temperatures $\tilde{\varepsilon}\rho_n \gg 1$, a $T \ll T_c^0$ (see calculations in appendixes A and B).

a) Close to superconductive transition temperature $(T_c^0 - T \ll T_c^0)$, $\tilde{\varepsilon}\,\rho_n \ll 1$, $\rho_n \ll 1$ for the functions $f_{mn}$ (m; n = 1, 2) on the basis of definitions (19) – (22) we obtain:

$$f_{11}(\rho_1,\tilde{\varepsilon}) = \frac{35}{24}\varsigma(3)\tilde{\varepsilon}\rho_1 - \frac{109\cdot 31}{40\cdot 16}\varsigma(5)\tilde{\varepsilon}^2\rho_1^2 + \cdots,$$

$$f_{12}(\rho_2) = \frac{7}{6}\varsigma(3)\rho_2 - \frac{31}{10}\varsigma(5)\rho_2^2 + \cdots,$$

$$f_{21}(\rho_1,\tilde{\varepsilon}) = \frac{7(3+2\tilde{\varepsilon})}{12(1+\tilde{\varepsilon})}\varsigma(3)\tilde{\varepsilon}\rho_1 - \frac{31(25+80\tilde{\varepsilon}+4\tilde{\varepsilon}^2)}{160(1+\tilde{\varepsilon})^2}\varsigma(5)\tilde{\varepsilon}^2\rho_1^2 + \cdots,$$

$$f_{22}(\rho_2) = \frac{7}{6}\varsigma(3)\rho_2 - \frac{31}{10}\varsigma(5)\rho_2^2 + \cdots,$$

(30)

where $\varsigma(x)$ - Rieman zeta-function.

b) Nearby zero temperature $T \ll T_c^0, \tilde{\varepsilon}\,\rho_n \gg 1$, the asymptotes of the $f_{mn}$ functions have the appearance:

$$f_{11}(\rho_1,\tilde{\varepsilon}) = \ln\left(\kappa\sqrt{\tilde{\varepsilon}\,\rho_1}\,\gamma\right) - \frac{1}{\pi^2\rho_1\tilde{\varepsilon}}\left[\varsigma'(2) - \frac{\varsigma(2)}{2}\ln\left(\frac{\tilde{\varepsilon}\,\rho_1\,\gamma\,\pi^2}{2e_0^{1/2}}\right)\right] + \cdots,$$

$$f_{12}(\rho_2,\tilde{\varepsilon}) = \ln\left(\frac{2\sqrt{2\rho_2}\,\gamma}{e_0}\right) - \frac{1}{\pi^2\rho_2}\left[\varsigma'(2) - \frac{\varsigma(2)}{2}\ln\left(\frac{\rho_2\,\gamma\,\pi\,(1+\tilde{\varepsilon})}{4}\right)\right] + \cdots,$$

(31)

$$f_{21}(\rho_1,\tilde{\varepsilon}) = \ln\left(c(\tilde{\varepsilon})\sqrt{\tilde{\varepsilon}\,\rho_1}\,\gamma\right) - \frac{1+\tilde{\varepsilon}}{2\pi^2\rho_1\tilde{\varepsilon}^{3/2}}\left[\varsigma'(2) - \frac{\varsigma(2)}{2}\left(\ln\left(\frac{\tilde{\varepsilon}\,\rho_1\,\gamma\,\pi^2}{1+\tilde{\varepsilon}}\right) - \frac{\tilde{\varepsilon}}{2}\right)\right] + \cdots,$$

$$f_{22}(\rho_2) = \ln\left(\frac{2\sqrt{2\rho_2}\,\gamma}{e_0}\right) - \frac{1}{\pi^2\rho_2}\left[\varsigma'(2) - \frac{\varsigma(2)}{2}\ln\left(\frac{\rho_2\,\gamma\,\pi^2}{2}\right)\right] + \cdots,$$

where $\kappa = \frac{1+\sqrt{2}}{\sqrt{2}}e_0^{1-\sqrt{2}} \approx 1.12$, $c(\tilde{\varepsilon}) = \frac{\sqrt{2.41(1+\sqrt{1+\tilde{\varepsilon}^2})}}{\sqrt{1+\tilde{\varepsilon}}}\exp(-0.207 + \frac{1}{2\tilde{\varepsilon}} - \frac{1}{2}\sqrt{1+\frac{1}{\tilde{\varepsilon}^2}})$, $e_0$ – base of the natural logarithm, $C = \ln\gamma$ is the Euler constant.

Let's pass to some simplifications assuming $\xi^{(1)}(T) = \xi^{(2)}(T) = \xi(T)$ and $\xi_1(T_c) = \xi_2(T_c^0) = \xi(T_c^0) = \xi_c^0$, then we'll consider every of the aforesaid temperature areas individually.

Case a) $\tilde{\varepsilon}\rho_n \ll 1$, $\rho_n \ll 1$, $T_c^0 - T \ll T_c^0$.

In this case we have dropped the terms, which contains products $f_{11}f_{22}$ and $f_{12}f_{21}$ because there are the coefficients $\lambda_{11}\lambda_{22}$, $\tilde{\lambda}_{12}\tilde{\lambda}_{21}$ between them, which are small for MgB$_2(\ll 1)$. In this case the equation will take form:

$$\lambda_{11}f_{11}(\rho_1,\tilde{\varepsilon}) + \lambda_{22}f_{22}(\rho_2) +$$

$$\xi(T_c^0)\left(\tilde{\lambda}_{12}\tilde{\lambda}_{21}f_{21}(\rho_1,\tilde{\varepsilon}) - \lambda_{11}\lambda_{22}f_{11}(\rho_1,\tilde{\varepsilon})\right) + f_{22}(\rho_2)\left[\lambda_{22} - a\xi(T_c)\right] - \nu\ln\left(\frac{T_c^0}{T}\right) = 0,$$

(32)

where $\nu = \sqrt{(\lambda_{11}-\lambda_{22})^2 + 4\tilde{\lambda}_{12}\tilde{\lambda}_{21}}$

Having substituted in (32) decompositions (31) and confining by the quadratic terms by the sought quantity $\rho_1 = \frac{\nu_1 \cdot eH_0}{(2\pi T)^2}$ we obtain the equation:

$$\tilde{\alpha}\,\rho_1^2 + \tilde{\beta}\,\rho_1 + \tilde{\chi} = 0$$

(33)

where

$$\tilde{\alpha} = -\frac{31}{10}\varsigma(5)\left[\frac{\lambda_{22}-a\xi_c}{\lambda^4}-\frac{\tilde{\varepsilon}^2}{16}\xi_c^0\left(\frac{109}{4}\lambda_{11}\lambda_{22}-\frac{(25+80\tilde{\varepsilon}+4\tilde{\varepsilon}^2)}{(1+\tilde{\varepsilon})^2}\tilde{\lambda}_{12}\tilde{\lambda}_{21}\right)+\frac{109}{64}\tilde{\varepsilon}^2\lambda_{11}\right],$$

$$\tilde{\beta} = \frac{7}{6}\varsigma(3)\left[\frac{\lambda_{22}-a\xi_c^0}{\lambda^2}-\frac{\tilde{\varepsilon}}{2}\xi_c\left(\frac{5}{2}\lambda_{11}\lambda_{22}-\frac{(3+2\tilde{\varepsilon})}{2(1+\tilde{\varepsilon})}\tilde{\lambda}_{12}\tilde{\lambda}_{21}\right)+\frac{5}{4}\tilde{\varepsilon}\lambda_{11}\right], \tag{34}$$

$$\tilde{\chi} = -\nu\left(\theta+\frac{\theta^2}{2}\right) \tag{35}$$

The solution of the equation (33) with the following decomposition of this solution by quantity $\theta = 1 - T/T_c^0$ gives us the behavior of the upper critical field $H_0(T) = H_{c2}^{(ab)0}(T)$ close to the superconductivity transition temperature:

$$\rho_c^0 \equiv \frac{\nu_1\,\nu_2\,e_0\,H_0(T)}{(2\pi T_c^0)^2} = \rho_1\frac{\nu_2}{\nu_1}\left(\frac{T}{T_c^0}\right)^{-2}\left(\frac{\nu\theta}{\tilde{\beta}}+\frac{\nu(\tilde{\beta}^2-2\tilde{\alpha}\,\nu)}{2\,\tilde{\beta}^3}\theta^2\right)\lambda^{-1}\left(\frac{T}{T_c^0}\right)^{-2},\ \lambda = \nu_1\Big/\nu_2 \tag{35}$$

b) The area of low temperatures $T \ll T_c$, $\tilde{\varepsilon}\rho_n \gg 1$

Substituting equations (25) into equation (24), it is not difficult to obtain the equation for determination of the upper critical field $H_0(T) = H_{c2}^0(T)$ in the area of low temperatures

$$a(\ln x)^2 + B\ln x + \bar{C} + F(T) = 0, \tag{36}$$

where

$$a = \lambda_{11}\lambda_{22}-\tilde{\lambda}_{12}\tilde{\lambda}_{21},\ B = \ln\left[\frac{2\sqrt{2}}{e_0}\right]a+\nu-\Lambda,$$

$$\bar{C} = \Lambda\xi_c^0-a\ln[\lambda]+\ln[\sqrt{\lambda}]\big(0.04a-a\ln[\sqrt{\tilde{\varepsilon}}]+\lambda_{11}-\lambda_{22}+\Lambda\big)+\ln[\kappa]\lambda_{11}- $$
$$-\big(0.4-a\ \ln[\sqrt{\tilde{\varepsilon}}]\big)\big(\Lambda+a\xi_c^0-\lambda_{22}\big), \tag{37}$$

$$\Lambda = \ln[c(\tilde{\varepsilon})]\tilde{\lambda}_{12}\tilde{\lambda}_{21}-\ln[\kappa]\lambda_{11}\lambda_{22},\ x = \sqrt{\tilde{\varepsilon}\gamma}\,\rho_c^0,\ \rho_c^0 = \frac{\nu_1\,\nu_2\,eH_0}{(2\pi T_c^0)^2}$$

At the same time $F(T)$ is defined by the expression:

$$F(T) = \tilde{\lambda}_{12}\tilde{\lambda}_{21}\left(\xi(T_c^0)-\ln\left[x\sqrt{\lambda}c(\tilde{\varepsilon})\right]+q_{21}\right)q_{12}-\left(\ln\left[\frac{2\sqrt{2}x}{e_0\sqrt{\lambda\tilde{\varepsilon}}}\right]-\xi(T_c^0)\right)q_{21}$$
$$-q_{22}\left(\left(\xi(T_c^0)-\ln\left[cx\sqrt{\lambda}\right]\right)\lambda_{11}-1\right)\lambda_{22}+q_{11}\lambda_{11}\left(\left(-\ln\frac{2\sqrt{2}x}{e_0\sqrt{\lambda\tilde{\varepsilon}}}+\xi(T_c^0)+q_{22}\right)\lambda_{22}-1\right) \tag{38}$$

where

$$q_{11}(T) = \frac{1}{\pi^2\rho_c^0\tilde{\varepsilon}\lambda}\left(\frac{T}{T_c^0}\right)^2\left[\varsigma'(2)-\frac{\varsigma(2)}{2}\ln\left(\frac{\tilde{\varepsilon}\rho_c^0\gamma\pi^2\lambda}{2e_0^{1/2}}\left(\frac{T_c^0}{T}\right)^2\right)\right],$$

$$q_{12}(T) = \frac{\lambda}{\pi^2\rho_c^0}\left(\frac{T}{T_c^0}\right)^2\left[\varsigma'(2)-\frac{\varsigma(2)}{2}\ln\left(\frac{\rho_c^0\gamma\pi^2(1+\tilde{\varepsilon})}{4\lambda}\left(\frac{T_c^0}{T}\right)^2\right)\right],$$

$$q_{21}(T) = \frac{(1+\tilde{\varepsilon})}{2\pi^2\rho_c^0\lambda\tilde{\varepsilon}^{3/2}}\left(\frac{T}{T_c^0}\right)^2\left[\varsigma'(2)-\frac{\varsigma(2)}{2}\left(\ln\left(\frac{\tilde{\varepsilon}^2\rho_c^0\gamma\pi^2\lambda}{1+\tilde{\varepsilon}}\left(\frac{T_c^0}{T}\right)^2\right)-\frac{\tilde{\varepsilon}}{2}\right)\right],$$

$$q_{22}(T) = \frac{\lambda}{\pi^2 \rho_c^0}\left(\frac{T}{T_c^0}\right)^2 \left[\varsigma'(2) - \frac{\varsigma(2)}{2}\ln\left(\frac{\rho_c^0 \gamma \pi^2}{2\lambda}\left(\frac{T_c^0}{T}\right)^2\right)\right].$$

The solution of the equation for pure MgB$_2$ at T = 0 is given by:

$$\rho_c^0 = \frac{v_1 v_2 e H_0}{(2\pi T_c^0)^2}\quad (\gamma\,\tilde{\varpi})^{-1}\quad \exp\left\{-\frac{B\pm\sqrt{B^2 - 4a\overline{C}}}{a}\right\}. \tag{38}$$

## 1. Case $\vec{H}\parallel c$. The determination of $H_{c2}^c$

We have described above the theory of the upper critical field for the two-band anisotropic system. In MgB$_2$ this value corresponds to the maximal value of the critical field. It is of interest to adduce the results of calculations of $H_{c2}^c$, which responds to the minimal value of this quantity in the intermetallic $MgB_2$ when the magnetic field $\vec{H}\parallel\vec{c}$. This information together with the behavior data adduced for $H_{c2}\parallel(ab)$ will allow obtaining the temperature dependence of the coefficient $\gamma_H = H_{c2}^{ab0}/H_{c2}^{c0}$, which determines the anisotropy for the upper critical field.

We consider $\vec{H}\parallel\vec{c}$. In this case it is possible to choose $A_x = A_z = 0$, $A_y = H_0(x+x')/2$ and we have for the phase multiplier:

$$2\varphi(\vec{r}',\vec{r}) = eH_0(x+x')(y'-y). \tag{39}$$

$H\parallel(ab)$ which was adduced in this work with taking into account of some significant differences, which are related to the peculiarity of the problem. For example in considered here case $\vec{H}\parallel\vec{c}$ the average velocity of the electrons in the plane (ab) plays important role for the both energetic bands and the meaning of the velocity directed along the axis z is found immaterial. This circumstance saves us from necessity of introducing the parameter, which defines the deviation of σ-band from two-dimensionality and makes the problem more anisotropic (see [27] for the details). We'll adduce some results. The critical magnetic field $H_{c2}^c$ is defined by the equation

$$a\tilde{f}_1(\rho_1)\tilde{f}_2(\rho_2) + \left[\lambda_{11} - a\xi_c^0\right]\tilde{f}_1(\rho_1) + \left[\lambda_{22} - a\xi_c^0\right]\tilde{f}_2(\rho_2) = 0, \tag{40}$$

where

$$\tilde{f}_1(\rho_1) = f_1(\rho_1) - \ln\left(\frac{T_c^0}{T}\right)^{1/2}\quad,\tilde{f}_2(\rho_2)\quad f_2(\rho_2) - \ln\frac{T_c^0}{T} \tag{41}$$

$$f_1(\rho_1) = \frac{\rho_1^{-1/2}}{\pi}\int_1^\infty \frac{du}{\sqrt{u^2-1}}\int_0^\infty \frac{d\xi}{\sinh\dfrac{\xi u}{\rho_1^{1/2}}}\left[1 - e^{\frac{-\xi^2 u^2}{2}}\right]. \tag{42}$$

Function $f_2(\rho_2)$ corresponds to the expression $f_{22}$ (22).

The asymptotes for the function $f_1(\rho_1)$ are given by:

$$f_1(\rho_1) = \frac{7}{8}\varsigma(3)\rho_1 - \frac{31\cdot 3}{32}\varsigma(5)\rho_1^2 + \cdots\quad \text{at } \rho_1 \ll 1,$$

$$f_1(\rho_1) = \frac{1}{4}\ln 2\gamma\rho_1 + \frac{1}{\rho_1}\frac{3}{2\pi^2}\varsigma(2) - \frac{1}{\rho_1^2}\frac{9}{4\pi^4}\varsigma(4) + \cdots \text{ at } \rho_1 \gg 1 \tag{43}$$

The asymptotes for the function $f_2(\rho_2)$ coincide with the formulas (30) and (32).

The solution of the equation (40) with taking into account the aforesaid asymptotes for the functions $f_n(\rho_n)$ allows us to obtain solution in the area of low temperature close to critical $(T_c - T) \ll T_c$, $\rho_n \ll 1$ and nearby zero temperature $(T \ll T_c)$, $\rho_n \gg 1$. These solutions, correspondingly, are given by:

At $\rho_n \ll 1$,

$$\rho_1(T) = \frac{v_1^2 e H_0(T)}{(2\pi T_{c0})^2} \quad \alpha_1 \theta + \alpha_2 \theta^2,$$ (44)

$$\alpha_1 = \frac{\eta_1 + 2\eta_2}{7\varsigma(3)\left[\dfrac{\eta_1}{4} + \dfrac{\eta_2}{3}\dfrac{1}{\lambda_1^2}\right]}, \qquad \alpha_2 = \alpha_1 \left\{\frac{\left[\dfrac{31\cdot 3}{16}\eta_1 + \dfrac{31}{5}\dfrac{1}{\lambda^2}\eta_2\right]\varsigma(5)\alpha_1}{\left[\dfrac{1}{4}\eta_1 + \dfrac{1}{3}\dfrac{1}{\lambda^2}\eta_2\right]7\varsigma(3)} + 1\right\}.$$ (45)

here

$$\theta = 1 - T\big/T_c^0$$

$$\eta_{1,2} = \frac{1}{2}\left[1 \pm \eta\right], \quad \eta \quad (\lambda_{11} - \lambda_{22})\left[(\lambda_{11} - \lambda_{22})^2 + 4\lambda_{11}\lambda_{22}\right]^{-1/2}.$$ (44)

At $\rho_n \gg 1$,

$$\rho_c^0(T) = \frac{v_1 v_2 e H_0(T)}{(2\pi T_c^0)^2} = \frac{e_0}{4\gamma} e^{\frac{1}{2}\eta - \frac{3}{2}\nu(1) + \Omega(\lambda)[1 - F_1^-(T)/\Omega(\lambda)]},$$ (45)

$$\eta_- = \frac{\lambda_{11} - \lambda_{22}}{a}, \quad \nu(1) \quad \frac{1}{a}\sqrt{(\lambda_{11} - \lambda_{22} - a\ln\lambda)^2 + 4\lambda_{12}\lambda_{21}},$$ (46)

$$\Omega(\lambda) \quad \left\{\left[=\ln\frac{e_0\lambda}{2} + \frac{3}{2}\eta - \frac{1}{2}\nu(1)\right]^2 + \frac{8\lambda_{12}\lambda_{21}}{a^2}\right\}$$ (47)

We determine $H_{c2}^{(c)}$ by the expression (44) close to the critical temperature $T_c - T \ll T_c$, and formula (45) defines the same quantity nearby zero temperature.

## 6. Application of the theory to MgB₂. Numeric computation and the discussion of the results. Anisotropy of the upper critical field

The behavior of the upper critical field H$_{c2}$ as a function of correlation $\dfrac{T}{T_c^0}$ was investigated in this work settling on the basis principles of the theory of superconductor in the magnetic field [22]-[24] on the whole temperature interval $0 < T < T_c^0$.

An anisotropic system equivalent to MgB₂ was considered in the parallel to the plane (ab) magnetic field: there were two overlapping energetic bands of the different dimension on the Fermi surface ($\sigma$-band was almost two-dimensional with a weak dispersion in the axis z direction and three-dimensional $\pi$-band for which the Fermi surface was chosen in the form of a sphere).

On the basis of the Ginzburg-Landau equations, generalizing the calculation methodic of Maki and Tsuzuki for the two-band case and taking into account aforesaid peculiarities of the band structure inherent in MgB₂, we obtain the system of equations for determination of $T_c^0$ (24), and also the upper critical field $H_{c2}^0 = H_{c2}^0(ab)$ (25) in the parallel to the plane (ab) magnetic field.

These equations can be used either the pure MgB₂ investigations or for more complex system obtained by partial substitution of Mg and B for other chemical elements.

Let's note that considered system of equations (24) and (25) doesn't contain dissipation of the electrons on the impurity potential, but takes into account just the energetic bands filling mechanism at introducing electrons or holes.

The asymptotes of the functions $f_{mn}$ (30), (32), which are engaged into the definition of the equation (25), were found and the analytic solutions were obtained close to the critical temperature $T_c^0 - T \ll T_c^0$, $(\tilde\varepsilon\rho_n \ll 1)$ (37) and also nearby zero temperature $T \ll T_c^0$, $(\tilde\varepsilon\rho_n \gg 1)$. The analytic solutions for a critical field $H_{c2}^{(c)}$, which corresponds to the direction of the parallel to the axis z magnetic field, were obtained earlier [28] and brief results were adduced in section 5 of this work.

There is a great dispersion by the value of the electron- phonon interaction parameters in the scientific literature. Therefore we adduced the numeric results using two sets of these parameters for MgB$_2$:

a) $\lambda_{11} = 0.362$; $\lambda_{22} = 0.172$; $\lambda_{12} = 0.054$; $\lambda_{21} = 0.053$ [29],

b) $\lambda_{11} = 0.302$; $\lambda_{22} = 0.135$; $\lambda_{12} = 0.04$; $\lambda_{21} = 0.038$ [17].

We take into account the inherent to MgB$_2$ correlation $\lambda_{11} > \lambda_{22} > \lambda_{12}, \lambda_{21}$.

The important role in the considerable difference $\rho_c^0(ab)$ from $\rho_c^0(c)$ is played by the strong anisotropy of the electronic energy spectrum, which contains weak dispersion of the electrons energy along the axis z (low values of the average velocities in this direction). In our theory this circumstance is defined by the presence of a small parameter $\tilde{\varepsilon}$. The small value of the interband electron-electron interaction also plays considerable role.

On Fig. 1 the dependencies $\rho_c^0(ab)$ and $\rho_c^0(c)$ on temperature $T$ at certain parameters of the theory are given.

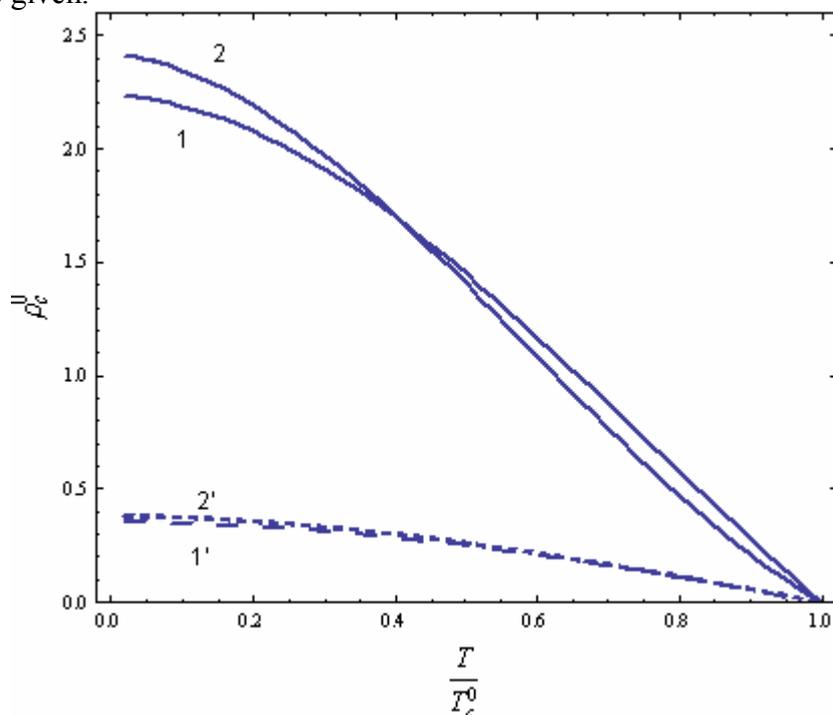

**Fig. 1**    *The dependence of the upper critical field $\rho_c$ on temperature at the values of theory parameters: curve 1 - $\tilde{\varepsilon} = 0.227$, $\lambda = v_1 / v_2 = 0.8$ and also the values of the renormalized constants of interaction $\lambda_{11} = 0.362$; $\lambda_{22} = 0.172$; $\lambda_{21} = 0.053$ ; $\lambda_{12} = 0.054$; Curve 2 - $\tilde{\varepsilon} = 0.117$, $\lambda = v_1 / v_2 = 0.8$, $\lambda_{11} = 0.362$; $\lambda_{22} = 0.255$; $\lambda_{12} = 0.174$; $\lambda_{21} = 0.165$. The dotted lines correspond to the values of quantity $\rho_c^0(c)$.*

On Fig. 2 it is adduced the behavior of the anisotropy coefficient $\gamma_H = \rho_c^0(ab) / \rho_c^0(c)$ depend on temperature. Continuous and dotted lines respond to the values of parameters on Fig. 1 correspondingly. The circles on this Figure respond to the experimental data of work Angst et al [30].

We have the qualitative picture of depending $\gamma_H$ on T at the theory and experiment comparison.

On Fig.3 it is adduced he dependence $\rho_c^0(ab)$ (curve 1) and $\rho_c^0(c)$ (curve 1′) for the case of set (b) $\tilde{\varepsilon} = 0.31$. Curve 2 contains small corrections in the definitions of parameters: $\lambda_{11} = 0.302$; $\lambda_{22} = 0.25$; $\lambda_{12} = 0.15$; $\lambda_{21} = 0.143$ at $\lambda = 0.8$; $\tilde{\varepsilon} = 0.133$. The dashed lines correspond to the field $\rho_c(c)$ with the same theory parameters.

On Fig. 4 it is presented the dependence of the anisotropy coefficient $\gamma_H$ on temperature.

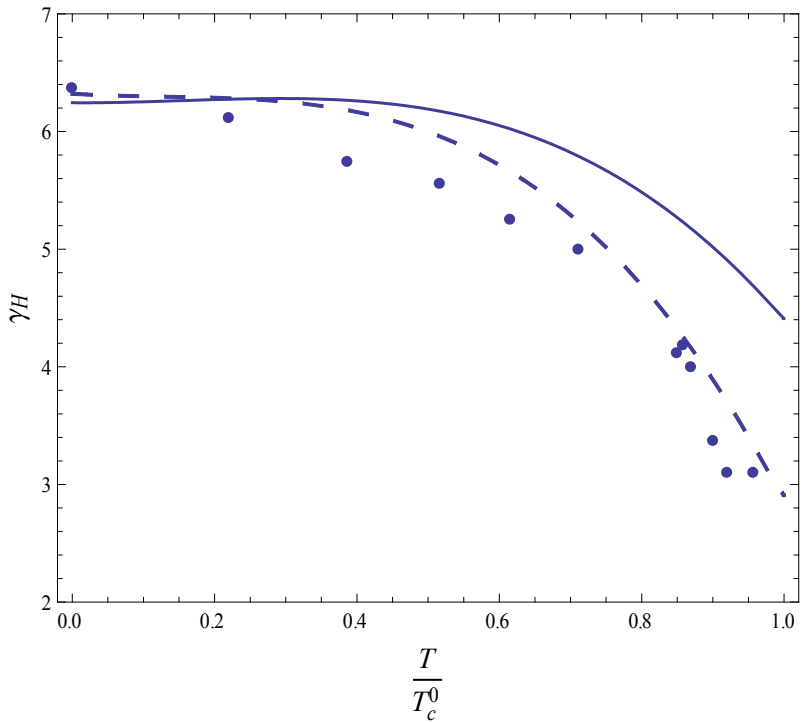

**Fig. 2** *The dependence of the anisotropy coefficient $\gamma_H = \rho_c^0(ab)/\rho_c^0(c)$ on temperature at the values of the theory parameters in accordance with continuous and dotted lines on Fig. 1. The circles on Fig.1 correspond to the experimental data of work [30].*

Curvatures 1, 2 are obtained on the basis of the same parameters of the theory that on Fig. 3. The circles on Fig. 4 correspond to the experimental data of work Lyard et al [31]. The comparison of the experimental points of the work [19], for example, with the data obtained in this work shows that there is the coincidence of the results. The same picture takes place in the theoretical dependencies of $\gamma_H$ on temperature.

The important result is the positive curvature of the upper critical field close to the temperature of superconductivity transition in the dependence of $\rho_c^0$ on $T$, which is typical for two-band systems.

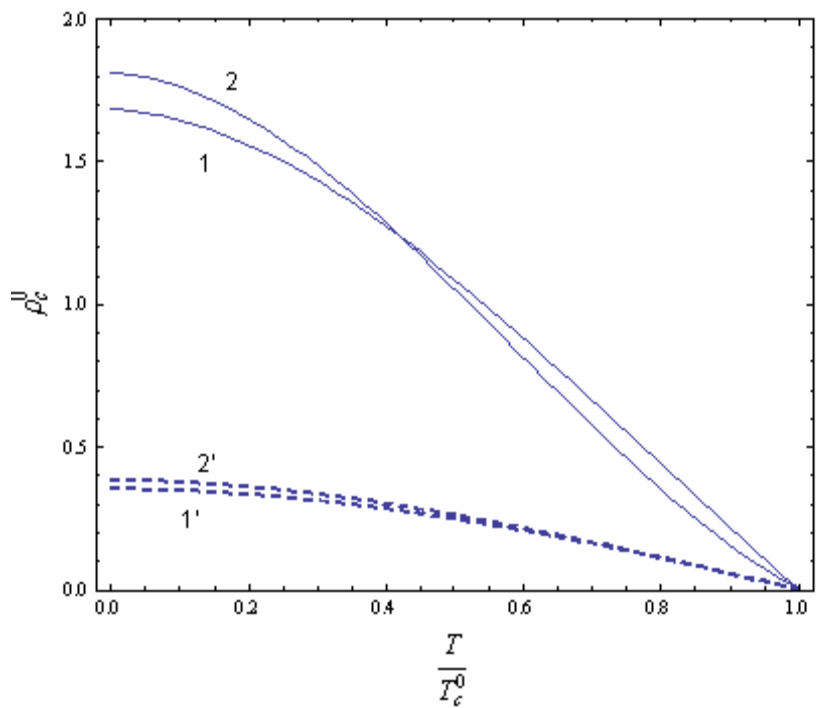

**Fig. 3** *The dependence of the dimensionless upper critical field $\rho_c^0$ on temperature for the values of the theory parameters: curve 1 - $\tilde{\varepsilon} = 0.31$, $\lambda = v_1 / v_2 = 0.8$, and also the values of the renormalized interaction constants $\lambda_{11} = 0.302; \lambda_{2\overline{2}} = 0.135; \lambda_{\overline{12}} = 0.04; \lambda_{\overline{21}} = 0.038$; curve 2 - $\tilde{\varepsilon} = 0.121$, $\lambda = 0.8$, $\lambda_{11} = 0.302; \lambda_{2\overline{2}} = 0.25; \lambda_{1\overline{2}} = 0.15; \lambda_{2\overline{1}} = 0.143$.*

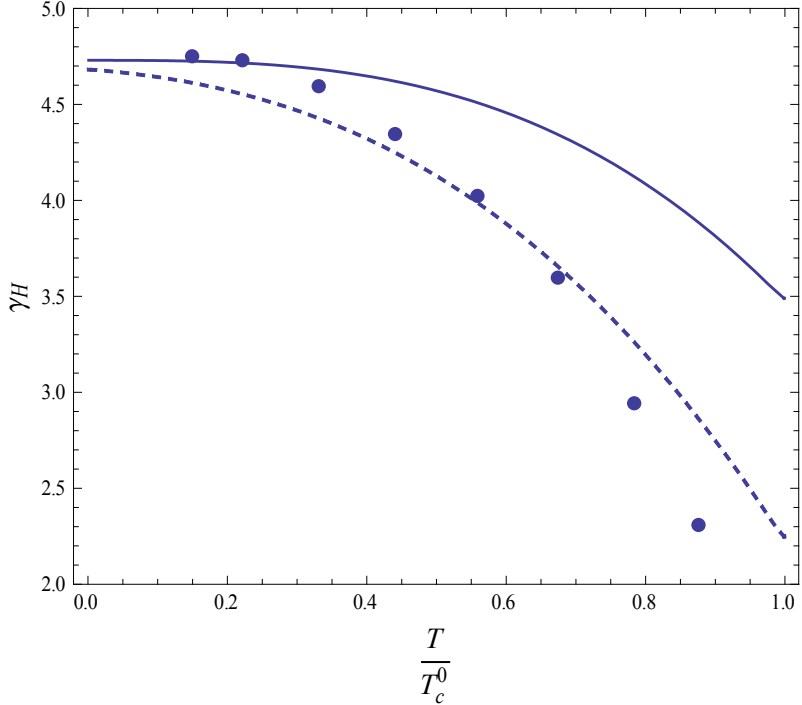

**Fig. 4** *The dependence of the anisotropy coefficient $\gamma_H = \rho_c^0(ab) / \rho_c^0(c)$ on temperature for the values of the theory parameters in accordance with continuous and dotted curves on Fig. 3 correspondingly. The circles on Fig.1 correspond to the experimental results of work [31].*

## Appendix A

We'll show here the transformations that pass from the equations (11), (12) to (15) and (16) correspondingly and to the definition of function $f_{mn}$.

After performing of integration in the equations (11) and (12) by the infinite limits, as it said after the formula (14), the system of equations (11) and (12) takes form:

$$\Delta_m^* = \sum_n \lambda_{mn} I_{mn} \Delta_n^*, \quad (m = 1, 2) \tag{A.1}$$

where

$$I_{11} = \sqrt{\frac{2\tilde{\varepsilon}eH_0}{\pi}} \frac{2T}{v_1} \int\limits_1^\infty \frac{du}{u\sqrt{u^2-1}} \int\limits_{-\infty}^{+\infty} dx \int\limits_{-\infty}^{+\infty} \frac{e^{-\tilde{\varepsilon}eH_0(x^2+x^2)}dx'}{\sinh\left(2\pi T(x-x')u/v_1\right)} \frac{\sin\left[u\tilde{\varepsilon}eH_0(x^2-x'^2)\right]}{\tilde{\varepsilon}eH_0(x^2-x'^2)} \theta\left(\left|x-x'\right|-\delta_{11}\right),$$

$$I_{12} = \sqrt{\frac{2\tilde{\varepsilon}eH_0}{\pi}} \frac{\pi T}{v_2} \int\limits_1^\infty \frac{du}{u} \int\limits_{-\infty}^{+\infty} dx \int\limits_{-\infty}^{+\infty} \frac{e^{-eH_0(x^2+\tilde{\varepsilon}x^2)}dx' J_0\left[(x^2-x'^2)eH_0\sqrt{u^2-1}\right]}{\sinh\left(2\pi T(x-x')u/v_2\right)} \theta\left(\left|x-x'\right|-\delta_{12}\right),$$

$$I_{21} = \sqrt{\frac{2eH_0}{\pi}} \frac{2T}{v_1} \int\limits_1^\infty \frac{du}{u\sqrt{u^2-1}} \int\limits_{-\infty}^{+\infty} dx \int\limits_{-\infty}^{+\infty} \frac{e^{-eH_0(x^2+\tilde{\varepsilon}x^2)}dx'}{\sinh\left(2\pi T(x-x')u/v_1\right)} \frac{\sin\left[u\tilde{\varepsilon}eH_0(x^2-x'^2)\right]}{\tilde{\varepsilon}eH_0(x^2-x'^2)} \times \tag{A.2}$$
$$\times \theta\left(\left|x-x'\right|-\delta_{21}\right),$$

$$I_{22} = \sqrt{\frac{2eH_0}{\pi}}\,\frac{\pi T}{v_2}\int_1^\infty \frac{du}{u}\int_{-\infty}^{+\infty}dx\int_{-\infty}^{+\infty}\frac{dx'\,e^{-eH_0(x^2+x'^2)}J_0\left[eH_0(x^2-x'^2)\sqrt{u^2-1}\right]}{\sinh\left(2\pi T(x-x')u/v_2\right)}\theta\left(|x-x'|-\delta_{22}\right),.$$

We'll introduce dimensionless quantities like $\xi = x(\tilde{\varepsilon}eH_0)^{1/2}$, $\xi' = x'(\tilde{\varepsilon}eH_0)^{1/2}$ for example in $I_{11}$, or $\xi = x(eH_0)^{1/2}$, $\xi' = x'(\tilde{\varepsilon}eH_0)^{1/2}$ in $I_{21}$, and then we'll obtain:

$$I_{11} = \sqrt{\frac{2}{\pi}}\,\frac{2T}{v_1}(\tilde{\varepsilon}eH_0)^{-1/2}\int_1^\infty \frac{du}{u\sqrt{u^2-1}}\int_{-\infty}^{+\infty}d\xi\int_{-\infty}^{+\infty}\frac{e^{-(\xi^2+\xi'^2)}d\xi'}{\sinh\left(2\pi T(\xi-\xi')u/(\tilde{\varepsilon}eH_0)^{1/2}v_1\right)}\times$$
$$\times\frac{\sin\left[u(\xi^2-\xi'^2)\right]}{(\xi^2-\xi'^2)}\theta\left(|\xi-\xi'|-\tilde{\delta}_{11}\right),$$

$$I_{12} = \sqrt{\frac{2}{\pi}}\,\frac{\pi T}{v_2}(eH_0)^{-1/2}\int_1^\infty \frac{du}{u}\int_{-\infty}^{+\infty}d\xi\int_{-\infty}^{+\infty}\frac{e^{-(\xi^2+\xi'^2)}d\xi'\,J_0\left[(\tilde{\varepsilon}^{-1}\xi^2-\xi'^2)\sqrt{u^2-1}\right]}{\sinh\left(2\pi T(\tilde{\varepsilon}^{-1/2}\xi-\xi')u/v_2(eH_0)^{1/2}\right)}\theta\left(|\tilde{\varepsilon}^{-1/2}\xi-\xi'|-\tilde{\delta}_{12}\right),$$

$$I_{21} = \sqrt{\frac{2}{\pi}}(eH_0)^{-1/2}\frac{2T}{v_1}\int_1^\infty \frac{du}{u\sqrt{u^2-1}}\int_{-\infty}^{+\infty}d\xi\int_{-\infty}^{+\infty}\frac{e^{-(\xi^2+\xi'^2)}d\xi'}{\sinh\left(2\pi T(\xi-\tilde{\varepsilon}^{-1/2}\xi')u/(eH_0)^{1/2}v_1\right)}\times$$
$$\times\frac{\sin\left[u\tilde{\varepsilon}(\xi^2-\tilde{\varepsilon}^{-1}\xi'^2)\right]}{\tilde{\varepsilon}(\xi^2-\tilde{\varepsilon}^{-1}\xi'^2)}\theta\left(|\xi-\tilde{\varepsilon}^{-1/2}\xi'|-\tilde{\delta}_{21}\right)$$

$$I_{22} = \sqrt{\frac{2}{\pi}}\,\frac{\pi T}{v_2}(eH_0)^{-1/2}\int_1^\infty \frac{du}{u}\int_{-\infty}^{+\infty}d\xi\int_{-\infty}^{+\infty}\frac{e^{-(\xi^2+\xi'^2)}d\xi'\,J_0\left[(\xi^2-\xi'^2)\sqrt{u^2-1}\right]}{\sinh\left(2\pi T(\xi-\xi')u/(eH_0)v_2\right)}\theta\left(|\xi-\xi'|-\tilde{\delta}_{22}\right),$$

$$(A.3)$$

where $\tilde{\delta}_{11} = (\tilde{\varepsilon}eH_0)^{1/2}\delta_{11}$, $\tilde{\delta}_{12} = (eH_0)^{1/2}\delta_{12}$, $\tilde{\delta}_{21} = (eH_0)^{1/2}\delta_{21}$, $\tilde{\delta}_{22} = (eH_0)^{1/2}\delta_{22}$.
Then the new variables are introduced:

$$\begin{matrix}\xi-\xi'=\varsigma \\ \xi+\xi'=2\eta\end{matrix}\ \text{in } I_{11},\quad \begin{matrix}\xi-\xi'=\varsigma \\ \xi+\xi'=2\eta\end{matrix}\ \text{in } I_{22},\quad \begin{matrix}\xi-\tilde{\varepsilon}^{-1/2}\xi'=\varsigma \\ \xi+\tilde{\varepsilon}^{-1/2}\xi'=2\eta\end{matrix}\ \text{in } I_{21},\quad \begin{matrix}\tilde{\varepsilon}^{-1/2}\xi-\xi'=\varsigma \\ \tilde{\varepsilon}^{-1/2}\xi+\xi'=2\eta\end{matrix}\ \text{in } I_{12}.\quad (A.4)$$

As a result of this $I_{mn}$ transforms to the following appearance:

$$I_{11} = (\tilde{\varepsilon}\rho_1)^{-1/2}\frac{2}{\pi}\sqrt{\frac{2}{\pi}}\int_1^\infty \frac{du}{u\sqrt{u^2-1}}\int_{\delta_{11}}^{+\infty}\frac{e^{\frac{-\varsigma^2}{2}}}{\sinh[\varsigma u(\tilde{\varepsilon}\rho_1)^{-1/2}]}d\varsigma\int_{-\infty}^{+\infty}e^{-2\eta^2}\frac{\sin[2\eta\varsigma u]}{2\eta\varsigma}d\eta,$$

$$I_{12} = \sqrt{\frac{2}{\pi}}\rho_2^{-1/2}\int_1^\infty \frac{du}{u}\int_{\delta_{12}}^{+\infty}\frac{\exp[\frac{-\varsigma^2(\tilde{\varepsilon}+1)}{4}]}{\sinh[\varsigma u\rho_2^{-1/2}]}d\varsigma\int_{-\infty}^{+\infty}\exp[-(\tilde{\varepsilon}-1)\varsigma\eta-(\tilde{\varepsilon}+1)\eta^2]J_0\left[2\eta\varsigma\sqrt{u^2-1}\right]d\eta,$$

$$I_{21} = \sqrt{\frac{2}{\pi}}\rho_1^{-1/2}\frac{2}{\pi}\int_1^\infty \frac{du}{u\sqrt{u^2-1}}\int_{\delta_{21}}^{+\infty}\frac{\exp[\frac{-\varsigma^2(\tilde{\varepsilon}+1)}{4}]}{\sinh[\varsigma u\rho_1^{-1/2}]}d\varsigma\int_{-\infty}^{+\infty}\exp[-(1-\tilde{\varepsilon})\varsigma\eta-(1+\tilde{\varepsilon})\eta^2]\frac{\sin[2\eta\varsigma xu]}{2\tilde{\varepsilon}\eta\varsigma}d\eta$$

$$I_{22} = \sqrt{\frac{2}{\pi}}\rho_2^{-1/2}\int_1^\infty \frac{du}{u}\int_{\delta_{22}}^{+\infty}\frac{e^{\frac{-\varsigma^2}{2}}}{\sinh[\varsigma u\rho_2^{-1/2}]}d\varsigma\int_{-\infty}^{+\infty}e^{-2\eta^2}J_0\left[2\eta\varsigma\sqrt{u^2-1}\right]d\eta,$$

$$(A.5)$$

We obtain after the integration by variable $\eta$ at last:

$$I_{11} = \xi^1 - f_{11} \ , I_{12} = \varepsilon^{-1/2}\sqrt{\frac{2}{\tilde{\varepsilon}+1}}\left(\xi^{(2)} - f_{12}\right), I_{21} = \sqrt{\frac{2}{\tilde{\varepsilon}+1}}\left(\xi^{(1)} - f_{21}\right), \ I_{22} = \xi^{(2)} - f_{22} \qquad (A.6)$$

where $f_{mn}$ are defined by formulas $(19) - (23)$ and $\xi^{(n)} = \xi_n(T)$ is defined by formula $(17)$.

At that

$$\xi^{(1)}(T) = \lim_{H_0 \to \infty}\left(\int_{-1}^{1} dy \int_{1}^{\infty}\frac{du}{u\sqrt{u^2-1}}\int_{\delta_{11}}^{+\infty}\frac{dx}{\sinh[xu]}\right) \qquad \xi^{(2)}(T) = \tilde{\varepsilon}^{1/2}\sqrt{\frac{2}{\tilde{\varepsilon}+1}}\lim_{H_0 \to \infty}\left(\int_{1}^{\infty}\frac{du}{u}\int_{\delta_{12}}^{+\infty}\frac{dx}{\sinh[xu]}\right)$$

$$\xi^{(1)}(T) = \sqrt{\frac{2}{\tilde{\varepsilon}+1}}\lim_{H_0 \to \infty}\left(\int_{-1}^{1} dy \int_{1}^{\infty}\frac{du}{u\sqrt{u^2-1}}\int_{\delta_{21}}^{+\infty}\frac{dx}{\sinh[xu]}\right) \quad \xi^{(2)}(T) = \lim_{H_0 \to \infty}\left(\int_{1}^{\infty}\frac{du}{u}\int_{\delta_{22}}^{+\infty}\frac{dx}{\sinh[xu]}\right)$$

$$(A.7)$$

The cut-off parameters $\delta_{ij}$ can be determined on the basis of correlations (A.7).

Let's introduce the renormalized interaction constants $\tilde{\lambda}_{21} = \sqrt{\frac{2}{\tilde{\varepsilon}+1}}\lambda_{21}$ and $\tilde{\lambda}_{12} = \tilde{\varepsilon}^{1/2}\sqrt{\frac{2}{\tilde{\varepsilon}+1}}\lambda_{12}$, then in accordance with the accepted notations the system of equations will transform to the following appearance:

$$\Delta_1^* = \lambda_{11}\Delta_1^*\xi^{(1)}(T) - \lambda_{11}f_{11}(\tilde{\varepsilon},\rho_1)\Delta_1^* + \tilde{\lambda}_{12}\xi^{(2)}(T)\Delta_2^* - \tilde{\lambda}_{12}f_{12}(\tilde{\varepsilon},\rho_2)\Delta_2^*$$
$$\Delta_2^* \ _{\cdots\cdots}\tilde{\lambda}_{21}\Delta_1^*\xi^{(1)}(\mp) - \tilde{\lambda}_{21}f_{21}(\tilde{\varepsilon},\rho_1)\Delta_1^* + \lambda_{22}\xi^{(2)}(T)\Delta_2^* - \lambda_{22}f_{22}(\varepsilon,\rho_2)\Delta_2^*$$

$$(A.8)$$

These equations can be rewritten in the following appearance:

$$\Delta_1^* = \lambda_{11}\Delta_1^*\xi^{(1)}(T_c) + \lambda_{11}\Delta_1^*\left[\xi^{(1)}(T) - \xi^{(1)}(T_c)\right] - \lambda_{11}f_{11}(\tilde{\varepsilon},\rho_1)\Delta_1^* +$$
$$+ \tilde{\lambda}_{12}\Delta_2^*\xi^{(2)}(T_c) + \tilde{\lambda}_{12}\left(\xi^{(2)}(T) - \xi^{(2)}(T_c)\right)\Delta_2^* - \tilde{\lambda}_{12}f_{12}(\tilde{\varepsilon},\rho_2)\Delta_2^*,$$
$$\Delta_2^* = \tilde{\lambda}_{21}\Delta_1^*\xi^{(2)}(T_c) + \tilde{\lambda}_{21}\Delta_1^*\left[\xi^{(1)}(T) - \xi^{(1)}(T_c)\right] - \tilde{\lambda}_{21}f_{21}(\tilde{\varepsilon},\rho_1)\Delta_1^* +$$
$$+ \lambda_{22}\Delta_2^*\xi^{(2)}(T_c) + \lambda_{22}\left(\xi^{(2)}(T) - \xi^{(2)}(T_c)\right)\Delta_2^* - \lambda_{22}f_{22}(\rho_2)\Delta_2^*$$

$$(A.9)$$

Equations (A.9) correspond to system of equations (15) and (16) of section 2.

## Appendix B

In this Appendix the method of the value $I_{11}$ calculations is adduced, the values $I_{12,}$ $I_{21}$, $I_{22}$ are calculated analogically.

We'll make the substitution of variable $\zeta = (\tilde{\varepsilon}\rho_1)^{1/2} x$ in $I_{11}$ (see A. 5).

$$I_{11} = \sqrt{\frac{2}{\pi}}\frac{2(\rho_1\tilde{\varepsilon})^{-1/2}}{\pi}\int_{1}^{\infty}\frac{du}{\sqrt{u^2-1}}\int_{\delta_{11}}^{\infty}d\zeta\, e^{-\frac{\zeta^2}{2}}\sinh^{-1}\left(\frac{\zeta u(2\pi T)}{(\tilde{\varepsilon}eH_0)^{-1/2}v_1}\right)\int_{-\infty}^{\infty}d\eta\, e^{-2\eta^2}\frac{\sin(2\eta\zeta u)}{2\eta\zeta u} =$$
$$= \sqrt{\frac{2}{\pi}}\frac{2}{\pi}\int_{1}^{\infty}\frac{du}{\sqrt{u^2-1}}\int_{\tilde{\delta}_{11}}^{\infty}dx\, e^{-\frac{\tilde{\varepsilon}\rho_1 x^2}{2}}\sinh^{-1}(xu)\int_{-\infty}^{\infty}d\eta\, e^{-2\eta^2}\frac{\sin(2\eta(\tilde{\varepsilon}\rho_1)^{1/2}xu)}{2\eta(\tilde{\varepsilon}\rho_1)^{1/2}xu},$$

$$(B.1)$$

Here $\tilde{\tilde{\delta}}_{11} = (\tilde{\varepsilon}\rho_1)^{-1/2}\tilde{\delta}_{11}$.

Let's consider the calculation of the asymptotic function $f_{11}$ in two limit cases:

  a) nearby zero temperature $(T_c^0 - T \ll T_c^0)$ and
  b) close to critical temperature $T \ll T_c^0$.

**a)** We'll introduce $\dfrac{\sin(2\eta\left(\tilde{\varepsilon}\rho_1\right)^{1/2}xu)}{2\eta\left(\tilde{\varepsilon}\rho_1\right)^{1/2}xu}$ in $I_{11}$ as $\dfrac{\sin(2\eta\left(\tilde{\varepsilon}\rho_1\right)^{1/2}xu)}{2\eta\left(\tilde{\varepsilon}\rho_1\right)^{1/2}xu}=\dfrac{1}{2}\int\limits_{-1}^{1}dy\,e^{i2\eta(\tilde{\varepsilon}\rho_1)^{1/2}xuy}$, after what

we'll make the integration in (B.1) by $\eta$ and using the following correlation

$$\frac{1}{\sinh(xu)}=\frac{1}{xu}+2xu\sum_{k=1}^{\infty}(-1)^k\frac{1}{(xu)^2+(\pi k)^2}\qquad \frac{1}{xu}+2xu\sum_{k=1}^{\infty}(-1)^k\int\limits_{0}^{\infty}dt\,e^{-(\pi k)^2 t-(xu)^2 t}$$

we'll introduce $I_{11}$ as a difference of 4 integral expressions:

$$\begin{aligned}I_{11}&=\xi^{(1)}-f_{11},\\ f_{11}&=f_{11}^{(1)}-f_{11}^{(2)}-f_{11}^{(3)},\end{aligned}\tag{B.2}$$

where

$$\begin{aligned}f_{11}^{(1)}&=\frac{1}{\pi}\int\limits_{-1}^{1}dy\int\limits_{1}^{\infty}\frac{du}{u\sqrt{u^2-1}}\int\limits_{\tilde{\delta}_{11}}^{+\infty}\frac{dx}{\sinh[xu]},\\ f_{11}^{(2)}&=\frac{1}{\pi}\int\limits_{-1}^{1}dy\int\limits_{1}^{\infty}\frac{du}{u\sqrt{u^2-1}}\int\limits_{\tilde{\delta}_{11}}^{\infty}\frac{dx}{x}e^{-\frac{\rho_2 x^2}{2}[1+u^2 y^2]},\\ f_{12}^{(3)}&=2\frac{1}{\pi}\sum_{k=1}^{\infty}(-1)^k\int\limits_{0}^{\infty}e^{-(\pi k)^2}dt\int\limits_{-1}^{1}dy\int\limits_{\tilde{\delta}_{11}}^{\infty}\frac{udu}{\sqrt{u^2-1}}\int\limits_{}^{\infty}x\cdot e^{-\frac{\rho_2 x^2}{2}[1+u^2 y^2]}dx\;,\end{aligned}\tag{B.3}$$

and $\xi^{(1)}$ is defined by formulae (17 and (A.7)).

Let's consider the calculation of every expression in (B.3) individually.

$$f_{11}^{(1)}=\frac{1}{\pi}\int\limits_{-1}^{1}dy\int\limits_{1}^{\infty}\frac{du}{u\sqrt{u^2-1}}\int\limits_{\tilde{\delta}_{11}}^{+\infty}\frac{dx}{\sinh[xu]}=\ln\left(\frac{1}{\tilde{\delta}_{11}}\right),\tag{B.4}$$

The $f_{11}^{(2)}$ will take the next form after integration by variable $x$ :

$$f_{11}^{(2)}=\frac{1}{\pi}\int\limits_{-1}^{1}dy\int\limits_{1}^{\infty}\frac{du}{u\sqrt{u^2-1}}\Gamma(0,\frac{\rho_1\tilde{\varepsilon}}{2}(1+(uy)^2))\,,$$

where $\Gamma(0,\dfrac{\rho_1\tilde{\varepsilon}}{2}(1+(uy)^2))$ is an incomplete gamma-function. We'll obtain after expressing this function through the integral exponential function $Ei(x)$ with its following asymptotic decomposing by $\tilde{\varepsilon}\rho\gg1$ the following:

$$f_{11}^{(2)}=-\frac{1}{\pi}\int\limits_{-1}^{1}dy\int\limits_{1}^{\infty}\frac{du}{u\sqrt{u^2-1}}\left(\ln\left(\frac{\gamma\tilde{\delta}_{11}^2}{2}\right)+\ln\left(1+(uy)^2\right)\right).$$

After the integration by u and y:

$$f_{11}^{(2)}=\ln\left(\frac{\sqrt{2}}{\sqrt{\gamma\left(1+\sqrt{2}\right)}\tilde{\delta}_{11}}\right)-(1-\sqrt{2})\,.\tag{B.5}$$

Let's pass now to the considering of $f_{11}^{(3)}$. We'll integrate it by $x$ and by u $f_{11}^{(3)}$, after that we'll obtain:

$$f_{11}^{(3)} = \frac{2}{\pi} \sum_{k=1}^{\infty} (-1)^k \int_0^{\infty} e^{-(\pi k)^2 t} dt \int_{-1}^{1} \frac{dy}{\sqrt{2t + \rho_1 \tilde{\varepsilon} y^2} \sqrt{(2t + \tilde{\varepsilon}\rho_1) + \tilde{\varepsilon}\rho_1 y^2}} = \tag{B.6}$$

$$= \frac{2}{\pi} \sum_{k=1}^{\infty} (-1)^k \int_0^{\infty} e^{-(\pi k)^2 t} dt \frac{1}{\tilde{\varepsilon}\rho_1} \left( \ln \frac{2(\tilde{\varepsilon}\rho_1)}{t} - \frac{1}{2} \right) + \cdots =$$

$$= -\frac{2}{\pi} \sum_{k=1}^{\infty} (-1)^k \frac{\left(\ln(2\tilde{\varepsilon}\rho_1 \gamma \pi^2 k^2) - 1/2\right)}{\tilde{\varepsilon}\rho_1 \pi^2} + \cdots \quad \frac{1}{\tilde{\varepsilon}\rho_1 \pi^2} \left( -\frac{\zeta(2)}{2} \ln \left[ \frac{\tilde{\varepsilon}\rho_1 \gamma \pi^2}{e_0^{1/2}} \right] + \zeta'(2) \right) + \cdots$$

We decomposed $\left(2t + \rho_1 \tilde{\varepsilon} y^2\right)^{-1/2} \left((2t + \tilde{\varepsilon}\rho_1) + \tilde{\varepsilon}\rho_1 y^2\right)^{-1/2}$ here by a small value $1/\tilde{\varepsilon}\rho_1$ having confined by the terms of order $\left(\tilde{\varepsilon}\rho_1\right)^{-1}$, then it was integrated by $y$ and by $t$ and summed by $k$. We'll obtain introducing formulas (B.4), (B.5), (B.6) in (B.2) at last:

$$f_{11}(\rho_1, \tilde{\varepsilon}) = \ln \left( \kappa \sqrt{\tilde{\varepsilon}\rho_1} \gamma \right) - \frac{1}{\pi^2 \rho_1 \tilde{\varepsilon}} \left[ \zeta'(2) - \frac{\zeta(2)}{2} \ln \left( \frac{\tilde{\varepsilon}\rho_1 \gamma \pi^2}{2e_0^{1/2}} \right) \right] + \cdots \tag{B.7}$$

$$\kappa = \frac{1 + \sqrt{2}}{\sqrt{2}} e_0^{1-\sqrt{2}}$$

**b)** Let's decompose $\sin(2\eta \left(\tilde{\varepsilon}\rho_1\right)^{1/2} xu)$ in Taylor series by a small quantity $\left(\tilde{\varepsilon}\rho_1\right)^{1/2}$ considering only the terms $\tilde{\varepsilon}\rho_1$ and $\left(\tilde{\varepsilon}\rho_1\right)^2$, and then we'll integrate it by $\eta$:

$$I_{11} = \frac{2}{\pi} \int_1^{\infty} \frac{du}{\sqrt{u^2 - 1}} \int_{\tilde{\delta}_{11}}^{\infty} e^{\frac{\tilde{\varepsilon}\rho_1 x^2}{2}} \sinh^{-1}(xu) \left( 1 - \frac{\tilde{\varepsilon}\rho_1 x^2 u^2}{6} + \frac{\left(\tilde{\varepsilon}\rho_1\right)^2 x^4 u^4}{40} \right) dx + \cdots = \tag{B.8}$$

$$= \xi^{(1)} - f_{11}$$

Here $f_{11} = \frac{2}{\pi} \int_1^{\infty} \frac{du}{\sqrt{u^2 - 1}} \int_{\tilde{\delta}_{11}}^{\infty} dx e^{-\frac{\tilde{\varepsilon}\rho_1 x^2}{2}} \sinh^{-1}(xu) \left( \frac{\tilde{\varepsilon}\rho_1 x^2 u^2}{6} - \frac{\left(\tilde{\varepsilon}\rho_1\right)^2 x^4 u^4}{40} \right) + \cdots. \tag{B.9}$

Let's decompose $e^{-\frac{\tilde{\varepsilon}\rho_1 x^2}{2}}$ by $\tilde{\varepsilon}\rho_1$

$$f_{11} = \frac{2}{\pi} \int_1^{\infty} \frac{du}{\sqrt{u^2 - 1}} \int_{\tilde{\delta}_{11}}^{\infty} dx \sinh^{-1}(xu) \left( \tilde{\varepsilon}\rho_1 x^2 \left( \frac{u^2}{6} + \frac{1}{2} \right) - \left(\tilde{\varepsilon}\rho_1\right)^2 x^4 \left( \frac{u^4}{40} + \frac{u^2}{12} + \frac{1}{8} \right) \right) + \cdots.$$

Then we'll obtain after integration by $x$:

$$f_{11} = \frac{2}{\pi} \int_1^{\infty} \frac{du}{\sqrt{u^2 - 1}} \left( \frac{7 \cdot \Gamma(3) \cdot \zeta(3) \cdot \tilde{\varepsilon}\rho_1}{4 \cdot u^3} \left( \frac{u^2}{6} + \frac{1}{2} \right) - \frac{31 \cdot \Gamma(5) \cdot \zeta(5) \left(\tilde{\varepsilon}\rho_1\right)^2}{64 \cdot u^5} \left( \frac{u^4}{40} + \frac{u^2}{12} + \frac{1}{8} \right) \right) + \cdots. \tag{B.10}$$

where $\Gamma(x)$ is a gamma-function.

The $f_{11}$ will take the next form after integration by $u$:

$$f_{11}(\rho_1, \tilde{\varepsilon}) = \frac{35}{24} \zeta(3) \tilde{\varepsilon}\rho_1 - \frac{109 \cdot 31}{40 \cdot 16} \zeta(5) \tilde{\varepsilon}^2 \rho_1^2 + \cdots \tag{B.11}$$